\documentstyle[12pt]{article}

\expandafter\ifx\csname mathrm\endcsname\relax\def\mathrm#1{{\rm #1}}\fi

\makeatletter
\if@twoside \oddsidemargin -17pt \evensidemargin 00pt
\else \oddsidemargin 00pt \evensidemargin 00pt
\fi
\expandafter\ifx\csname mathrm\endcsname\relax\def\mathrm#1{{\rm #1}}\fi

\makeatother

\def\beq{\begin{equation}}
\def\eeq{\end{equation}}
\def\beqar{\begin{eqnarray}}
\def\eeqar{\end{eqnarray}}
\def\barr#1{\begin{array}{#1}}
\def\earr{\end{array}}
\def\bfi{\begin{figure}}
\def\efi{\end{figure}}
\def\btab{\begin{table}}
\def\etab{\end{table}}
\def\bce{\begin{center}}
\def\ece{\end{center}}
\def\nn{\nonumber}

\def\text{\textstyle}

\def\ga{\gamma}
\def\de{\delta}

\def\si{\sigma}

\def\Ga{\Gamma}
\def\De{\Delta}

\def\refeq#1{\mbox{(\ref{#1})}}
\def\reffi#1{\mbox{Fig.~\ref{#1}}}

\def\refta#1{\mbox{Tab.~\ref{#1}}}
\def\refse#1{\mbox{Sect.~\ref{#1}}}

\def\citere#1{\mbox{Ref.~\cite{#1}}}
\def\citeres#1{\mbox{Refs.~\cite{#1}}}

\def\solid{\raise.9mm\hbox{\protect\rule{1.1cm}{.2mm}}}
\def\dash{\raise.9mm\hbox{\protect\rule{2mm}{.2mm}}\hspace*{1mm}}

\newcommand{\GeV}{\unskip\,\mathrm{GeV}}
\newcommand{\MeV}{\unskip\,\mathrm{MeV}}
\newcommand{\TeV}{\unskip\,\mathrm{TeV}}
\newcommand{\fba}{\unskip\,\mathrm{fb}}

\newcommand{\ord}{{\cal O}}

\def\mathswitchr#1{\relax\ifmmode{\mathrm{#1}}\else$\mathrm{#1}$\fi}

\newcommand{\PV}{\mathswitchr V}
\newcommand{\PW}{\mathswitchr W}
\newcommand{\PZ}{\mathswitchr Z}

\newcommand{\PH}{\mathswitchr H}
\newcommand{\Pe}{\mathswitchr e}

\newcommand{\Pt}{\mathswitchr t}

\newcommand{\Pep}{\mathswitchr {e^+}}
\newcommand{\Pem}{\mathswitchr {e^-}}

\newcommand{\PWp}{\mathswitchr {W^+}}
\newcommand{\PWm}{\mathswitchr {W^-}}

\def\mathswitch#1{\relax\ifmmode#1\else$#1$\fi}

\newcommand{\Mf}{\mathswitch {m_f}}

\newcommand{\MV}{\mathswitch {M_\PV}}
\newcommand{\MW}{\mathswitch {M_\PW}}

\newcommand{\MZ}{\mathswitch {M_\PZ}}
\newcommand{\MH}{\mathswitch {M_\PH}}
\newcommand{\Me}{\mathswitch {m_\Pe}}

\newcommand{\Mt}{\mathswitch {m_\Pt}}

\newcommand{\scrs}{\scriptscriptstyle}
\newcommand{\sw}{\mathswitch {s_{\scrs\PW}}}
\newcommand{\cw}{\mathswitch {c_{\scrs\PW}}}

\newcommand{\GF}{\mathswitch {G_\mu}}

\def\Re{\mathop{\mathrm{Re}}\nolimits}
\def\Im{\mathop{\mathrm{Im}}\nolimits}

\def\rhow{\rho_{\scrs\PW}}

\def\draftdate{\relax}
\def\mda{\relax}
\def\mua{\relax}
\def\mla{\relax}
\def\draft{
\def\thtystars{******************************}
\def\sixtystars{\thtystars\thtystars}
\typeout{}
\typeout{\sixtystars**}
\typeout{* Draft mode!
         For final version remove \protect\draft\space in source file *}
\typeout{\sixtystars**}
\typeout{}
\def\draftdate{\today}
\def\mua{\marginpar[\boldmath\hfil$\uparrow$]%
                   {\boldmath$\uparrow$\hfil}%
                    \typeout{marginpar: $\uparrow$}\ignorespaces}
\def\mda{\marginpar[\boldmath\hfil$\downarrow$]%
                   {\boldmath$\downarrow$\hfil}%
                    \typeout{marginpar: $\downarrow$}\ignorespaces}
\def\mla{\marginpar[\boldmath\hfil$\rightarrow$]%
                   {\boldmath$\leftarrow $\hfil}%
                    \typeout{marginpar: $\leftrightarrow$}\ignorespaces}
\def\Mua{\marginpar[\boldmath\hfil$\Uparrow$]%
                   {\boldmath$\Uparrow$\hfil}%
                    \typeout{marginpar: $\Uparrow$}\ignorespaces}
\def\Mda{\marginpar[\boldmath\hfil$\Downarrow$]%
                   {\boldmath$\Downarrow$\hfil}%
                    \typeout{marginpar: $\Downarrow$}\ignorespaces}
\def\Mla{\marginpar[\boldmath\hfil$\Rightarrow$]%
                   {\boldmath$\Leftarrow $\hfil}%
                    \typeout{marginpar: $\Leftrightarrow$}\ignorespaces}
\overfullrule 5pt
\oddsidemargin -15mm
\marginparwidth 29mm
}

\makeatletter

\def\eqnarray{\stepcounter{equation}\let\@currentlabel=\theequation
\global\@eqnswtrue
\global\@eqcnt\z@\tabskip\@centering\let\\=\@eqncr
$$\halign to \displaywidth\bgroup\hskip\@centering
  $\displaystyle\tabskip\z@{##}$\@eqnsel&\global\@eqcnt\@ne
  \hskip 2\arraycolsep \hfil${##}$\hfil
  &\global\@eqcnt\tw@ \hskip 2\arraycolsep $\displaystyle\tabskip\z@{##}$\hfil
   \tabskip\@centering&\llap{##}\tabskip\z@\cr}
\def\appendix{\par
 \setcounter{section}{0} \setcounter{subsection}{0}
 \def\thesection{\Alph{section}}}

\makeatother

\newcommand{\lsim}
{\;\raisebox{-.3em}{$\stackrel{\displaystyle <}{\sim}$}\;}
\newcommand{\gsim}
{\;\raisebox{-.3em}{$\stackrel{\displaystyle >}{\sim}$}\;}

\begin{document}
\thispagestyle{empty}
\def\thefootnote{\fnsymbol{footnote}}
\setcounter{footnote}{1}
\null
\hfill BI-TP 96/45 \\
\null
\hfill hep-ph/9610529
\vskip .8cm
\begin{center}
{\Large \bf \boldmath{
Theoretical aspects of W-pair production in $\Pep\Pem$ collisions}}%
\footnote{Talk given at the {\it 3$\,^{rd}$ International Symposium on
Radiative Corrections}, Cracow, Poland, August 1-5, 1996.}
\vskip 3em
{\large Stefan Dittmaier}
\vskip .5em
{\it Fakult\"at f\"ur Physik, Universit\"at Bielefeld, \\
Postfach 10 01 31, D-33501 Bielefeld, Germany}
\end{center} \par
\vskip 2.0cm
\vfil
{\bf Abstract} \par
The most interesting theoretical features of W-pair production in
$\Pep\Pem$ collisions are reviewed. Based on an analysis of on-shell
W-pairs, it is shown that the mere inclusion of leading 
${\cal O}(\alpha)$ corrections---such as initial-state radiation,
effects from the universal quantities $\De\alpha$ and $\De r$, and
the Coulomb singularity---cannot be expected to approximate the 
${\cal O}(\alpha)$-corrected cross-section of $\Pep\Pem\to\PW\PW\to 4f$
to better than 1-2\% for LEP2 energies. The importance of the
gauge-invariance problem, arising
from the introduction of finite decay widths, is discussed. The simplest
consistent method to include ${\cal O}(\alpha)$ corrections at least to
the doubly resonant contributions, 
viz.\ a pole expansion around
the double resonance, is also briefly described.
\par
\vskip .8cm
October 1996
\null

\setcounter{page}{0}
\clearpage
\def\thefootnote{\arabic{footnote}}
\setcounter{footnote}{0}

\section{Introduction}

\begin{sloppypar}
LEP2 provides first experimental results on the process
$\Pep\Pem\to\PW\PW\to4f$ at center-of-mass (CMS) energies between
$161\GeV$ and roughly $200\GeV$. 
Future experiments at the next linear collider (NLC), which is planned 
for CMS energies of 0.5-2$\TeV$, will be able to explore higher energies.
\end{sloppypar}

The most prominent goal to be reached via W-pair production at LEP2 is a
precise determination of the W-boson mass $\MW$. A reduction of the
present experimental uncertainty of $\Delta\MW=125\MeV$~\cite{MW} 
by a factor of 3-4 is aimed at. The most promising methods for the
extraction of $\MW$ from the data are the investigation of the total
cross-section near threshold and the reconstruction of $\MW$ from the
W-decay products, as extensively discussed in \citere{CERN9601wmass}.
For the first method the total cross-section at $\sqrt{s}=161\GeV$, where
the maximal sensitivity to $\MW$ is reached, has to be known with an
theoretical accuracy of 1-2\% (half of the expected statistical
error). The direct reconstruction relies on the invariant-mass
distribution of the decay products which is strongly
influenced by effects of initial-state radiation.
Consequently, the inclusion of radiative corrections (RCs) in theoretical
predictions is indispensable for both methods.

The second important issue accessible by $\Pep\Pem\to\PW\PW$ is the
investigation of the triple-gauge-boson couplings. Although the
non-abelian self-couplings of the Standard Model could be indirectly
confirmed by the LEP1 observables and the W-boson mass via radiative 
corrections~\cite{LEP1rcs}, only rather crude experimental bounds on the 
general (anomalous) structure of bosonic self-couplings exist, which 
result from hadron collisions. Although $\Pep\Pem\to\PW\PW$ is the first
observable $\Pep\Pem$-process, where non-abelian self-couplings enter
already lowest-order predictions, the sensitivity of the 
total cross-section
to anomalous couplings at LEP2 energies is too weak to yield stringent
bounds, and one is forced to inspect the angular distributions of the
produced W bosons (see e.g.\ \citere{CERN9601ancoup}). Since also
RCs distort angular distributions, these mimic
anomalous couplings and therefore have to be extracted. At NLC energies
much more drastic effects of anomalous couplings can arise, but also
RCs become more important.

The above considerations underline the importance of RCs
to W-pair production, which are the 
subject of this short
review. More detailed reviews can be found in
\citeres{Be94,CERN9601wwcross}. Following two steps of
sophistication, we consider on-shell W-pairs in \refse{se:onshell} before
dealing with the complete four-fermion production process in
\refse{se:4fermion}. Section~\ref{se:concl} contains our conclusions.

\section{On-shell W-pair production}
\label{se:onshell}

\subsection{Born cross-section}

We consider the reaction
\beq
\Pep(p_+,\kappa_+) + \Pem(p_-,\kappa_-) \;\to\;
\PWp(k_+,\lambda_+) + \PWm(k_-,\lambda_-),
\eeq
where $p_\pm$, $k_\pm$ and $\kappa_\pm$, $\lambda_\pm$ denote the
momenta and helicities of the respective particles. The Mandelstam
variables are defined by
\beq
s=4E^2, \qquad t=\MW^2-2E^2(1-\beta\cos\theta),
\eeq
where $E$ is the beam energy, 
$\beta=\sqrt{1-\MW^2/E^2}$ the W-boson velocity, and
$\theta$ the scattering angle in the CMS. 
\begin{figure}[b]
\begin{center}
\begin{picture}(10.5,2.0)
\put(-5.5,-23.2){\includegraphics{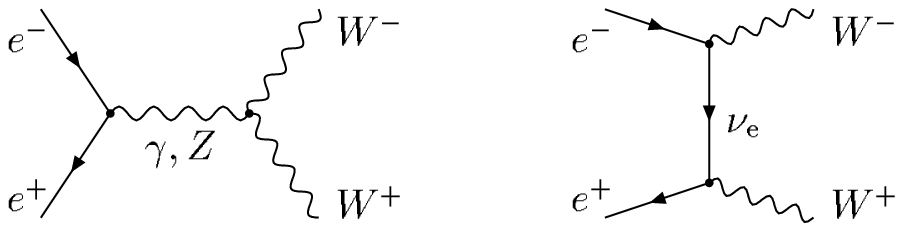}}
\end{picture}
\end{center}
\caption{Born diagrams for \protect{$\Pep\Pem\to\PWp\PWm$}.}
\label{fig:trdiag}
\efi
The lowest-order amplitude ${\cal M}^\kappa_{\mathrm{B}}$ gets 
contributions from the diagrams shown in \reffi{fig:trdiag} and reads
\beq
{\cal M}^\kappa_{\mathrm{B}} = 
\frac{e^2}{2\sw^2}\frac{1}{t} {\cal M}^\kappa_t \delta_{\kappa-} +
e^2\biggl[\frac{1}{s}-\biggl(1-\frac{1}{2\sw^2}\delta_{\kappa-}\biggr)
\frac{1}{s-\MZ^2}\biggr]{\cal M}^\kappa_s,
\eeq
where $\kappa=\pm$ is used as a shorthand for the electron helicity,
which is conserved $(\kappa=\kappa_-=-\kappa_+)$ since the electron mass
$\Me$ is neglected whenever possible. 
We calculate in the on-shell scheme where the weak mixing angle is fixed
by the ratio of gauge-boson masses, $\cw^2 = 1-\sw^2 = \MW^2/\MZ^2$.
The quantities ${\cal M}^\kappa_t$
and ${\cal M}^\kappa_s$ for the $t$- and $s$-channel graphs, respectively,
contain the spinors and polarization vectors 
and depend on $s$, $t$, $\kappa$, and $\lambda_\pm$.

Near threshold ($\beta\to0$) the $s$-channel contribution is suppressed by 
a factor $\beta$ so that the $t$-channel contribution 
dominates, and the differential cross-section reads
\beq
\frac{d\sigma_{\mathrm{B}}}{d\Omega} = 
\frac{\alpha^2}{4\sw^4}\frac{\beta}{s}
\biggl[1+4\beta\cos\theta\frac{3\cw^2-1}{4\cw^2-1}+{\cal O}(\beta^2)\biggr].
\eeq
In the high-energy limit ($E\to\infty$) the total cross-section behaves as
\beq
\sigma_{\mathrm{B}} = 
\frac{\pi\alpha^2}{4\sw^4}\frac{1}{s}
\biggl[2\log\biggl(\frac{s}{\MW^2}\biggr)-\frac{5}{2}-\frac{1}{3\cw^2}
+\frac{5}{24\cw^4}+{\cal O}(\log s/s)\biggr].
\eeq
The presence of gauge cancellations between $t$- and $s$-channel contributions 
is necessary to guarantee unitarity for longitudinally polarized W bosons,
i.e.\ the leading behavior of ${\cal M}^\kappa_t$ and ${\cal M}^\kappa_s$
is correlated. However, there is of course no high-energy relation
between the contributions originating from the different gauge
couplings, which lead to the following decomposition of  
${\cal M}^\kappa_{\mathrm{B}}$:
\beq
\label{eq:MB}
{\cal M}^\kappa_{\mathrm{B}} = 
\frac{e^2}{2\sw^2} {\cal M}^\kappa_I \delta_{\kappa-} +
e^2 {\cal M}^\kappa_Q.
\eeq
${\cal M}^\kappa_I$ defines the ``isospin part'' with the associated
coupling $e/\sw$, and ${\cal M}^\kappa_Q$ the ``electromagnetic 
part'' with the coupling $e$.

\subsection{Radiative corrections}

The ${\cal O}(\alpha)$ RCs to the
W-pair production cross-section consist of three different parts: 
\beq
d\sigma = d\sigma_{\mathrm{B}} 
(1+\delta_{\mathrm{V}}+\delta_{\mathrm{S}}+\delta_{\mathrm{H}}),
\eeq
the virtual one-loop 
correction $\delta_{\mathrm{V}}$, the real soft-photonic correction
$\delta_{\mathrm{S}}$, and the hard bremsstrahlung correction
$\delta_{\mathrm{H}}$. 
Both the complete virtual~\cite{eewwvirt} and real~\cite{eewwreal} 
corrections were calculated by different groups.

The lowest-order and ${\cal O}(\alpha)$-corrected total cross-sections
are shown in \reffi{fig:intcs}, where here and in the following
$\si_{\mathrm{B}}$ is parametrized by $\alpha$, $\MZ$, $\MW$, and the
input parameters are taken from \citere{CERN9601wwcross}.
The size of the relative RCs $\de$, which is also depicted in
\reffi{fig:intcs}, is at the order of 5-10\% for LEP2 energies.
The large negative amount of $\de$ near threshold indicates the
necessity of soft-photon exponentiation in this region.
For NLC energies the RCs are at the order of 20\%, but we note that the
RCs are further enhanced if the forward direction is excluded by an
angular cut.
\begin{figure}
\begin{center}
\begin{picture}(12.5,13.2)
\put(-1.9,-10.6){\includegraphics{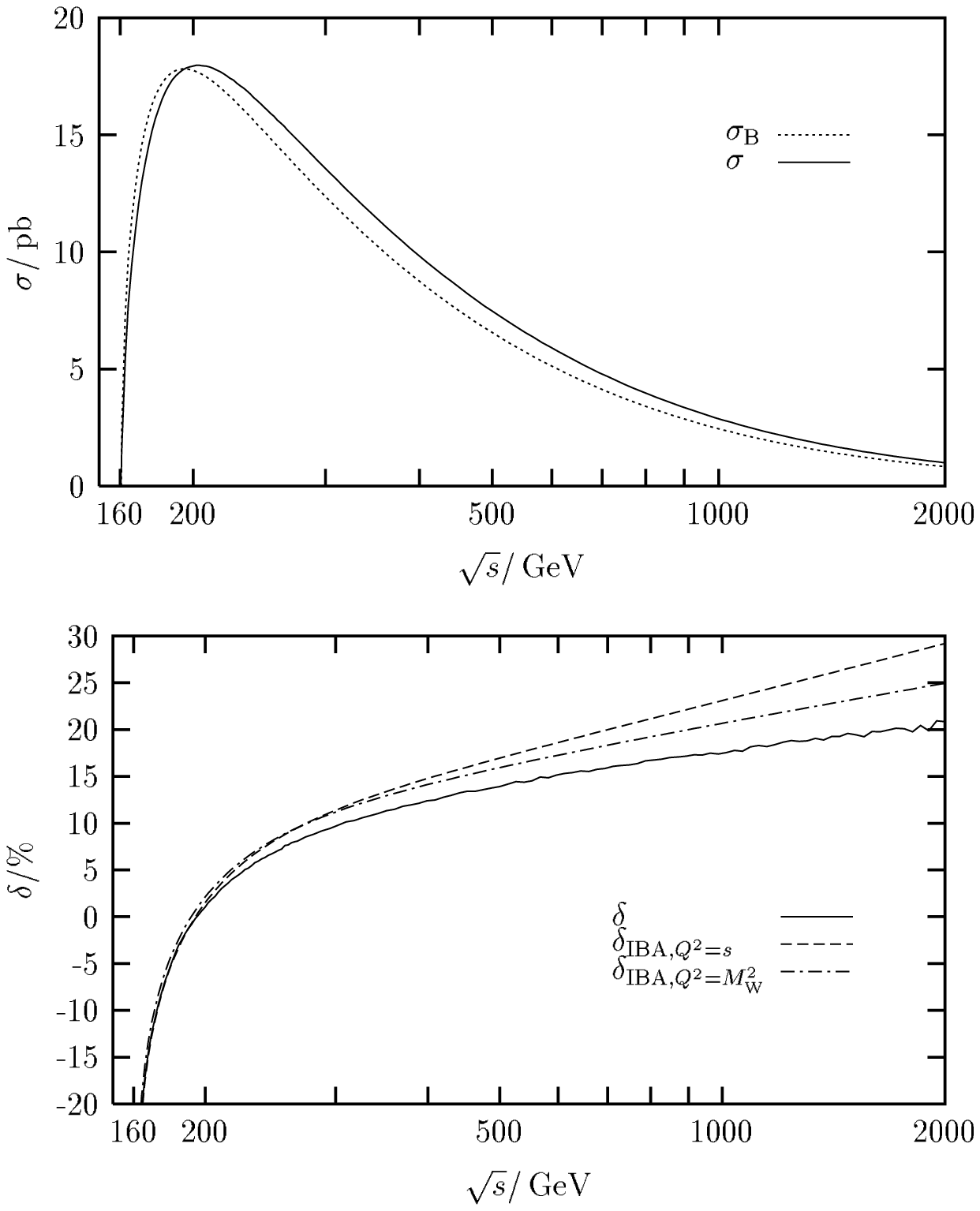}}
\end{picture}
\end{center}
\caption{
Born cross-section $\sigma_{\mathrm{B}}$, 
$\ord(\alpha)$-corrected cross-section (including also leading
higher-order effects via $\De\alpha$ and $\De r$) 
$\sigma=\sigma_{\mathrm{B}}(1+\de)$, and relative
correction $\de$ for unpolarized particles. 
The improved Born approximation (IBA) of \refeq{eq:IBA} is
also shown for two different choices of the photon splitting scale
$Q^2$.
}
\label{fig:intcs}
\efi

Concerning RCs the differences
between W-pair and fermion-pair production at LEP2 and LEP1,
respectively, are characterized by two features: Firstly, there is no
simple effective Born approximation for W-pair production, and thus 
the RCs cannot be absorbed into effective couplings. Secondly, there is no
gauge-invariant, unambiguous splitting between QED and genuinely weak
corrections for W pairs. Moreover, the analytical structure of the RCs is
involved resulting in long and rather slow computer codes.

\begin{sloppypar}
A first step 
towards a successful approximation of the RCs is 
to control the leading corrections, which are of the following
origin:
\renewcommand{\labelenumi}{(\roman{enumi})}
\begin{enumerate}
\item Initial-state radiation: \\
Collinear initial-state radiation (ISR) off the incoming electrons and
positrons yield large corrections of $\ord(\alpha^n\log^n\Me^2/Q^2)$ with
$n=1,2,\dots$. The leading logarithms are universal and can be most
easily calculated via the structure-function approach~\cite{be89}:
\beq
\label{eq:ISR}
\delta\sigma_{\mathrm{ISR}}(s) =
\int_{4\MW^2/s}^1 dx\,\Phi(x,Q^2)\sigma_{\mathrm{B}}(xs),
\eeq
i.e.\ by a convolution of the lowest-order cross-section with a flux
function $\Phi$. In $\ord(\alpha)$ the flux function reads 
\beqar
\Phi(x,Q^2) &=& \frac{\alpha}{\pi}\log\biggl(\frac{Q^2}{\Me^2}\biggr)
\biggl[\delta(1-x)
\biggl(\frac{3}{2} +2\log\frac{\Delta E}{E}\biggr)
\nn\\ && \quad {}
+\theta\biggl(1-\log\frac{\Delta E}{E}-x\biggr)
\frac{1+x^2}{1-x}\biggr]
+\ord(\alpha^2),
\hspace{2em}
\eeqar
but $\Phi$ is also known up to $\ord(\alpha^3\log^3\Me^2/Q^2)$ 
(see \citere{CERN9601wwcross} and references therein). 
The soft-photon cutoff $\Delta E$, which drops
out in the final result, separates the contributions of soft
($E_\gamma<\Delta E$) and hard ($E_\gamma>\Delta E$) photons. The
splitting scale $Q^2$ is not determined in leading logarithmic
approximation and thus expresses part of the theoretical uncertainty
owing to neglected non-leading corrections. At LEP2 energies ISR
corrections amount to roughly 6\%.
\item Running of the electromagnetic coupling $\alpha(q^2)$: \\
The definition of the $\Pe\Pe\gamma$ coupling $e$ in the low-energy limit
(Thomson scattering) leads to large universal corrections to processes
which involve energies at the electroweak scale. These corrections,
which are of $\ord(\alpha^n\log^n\Mf^2/q^2)$ 
for light fermions $f$, can be accounted for
by introducing the running electromagnetic coupling
$\alpha(q^2)$, which replaces the fine-structure constant
$\alpha(0)=e^2/4\pi=1/137.0\dots$ as input parameter. In perturbation
theory $\alpha(q^2)$ is parametrized by $\alpha(0)$ and the light
fermion masses $\Mf$
where the quark masses are adjusted to reproduce the 
hadronic vacuum polarization~\cite{bu95}. At LEP2 energies the
transition from $\alpha(0)$ to $\alpha(q^2)$ 
corrects the cross-section by about 8\%.
\item Coulomb singularity: \\
Near threshold the long range of the Coulomb interaction between slowly
moving W~bosons causes large universal corrections, which are
proportional to the Born cross-section,
\beq
\delta\sigma_{\mathrm{Coul}} 
\;\raisebox{-.3em}{$\widetilde{\;\scriptstyle\beta\to0\;}$}\;
\frac{\alpha\pi}{2\beta}\sigma_{\mathrm{B}}.
\eeq
The singular behavior of the relative correction at $\beta\to0$ is a
peculiarity of the on-shell approximation, which assumes stable
W~bosons. This singularity is regularized by a finite W~decay width,
since the range of interaction is effectively truncated by the W~decay
(see next section).
\item Effects from light or heavy masses: \\
In analogy to the long-range effect of the Coulomb interaction, which is
due to the exchange of massless photons, the exchange of a relatively
light Higgs boson between slowly moving W~bosons near threshold leads to
a Yukawa-like interaction. In the limit $\MH\ll\MW$ the leading behavior
is given by 
\beq
\delta\sigma_{\mathrm{LH}} 
\;\raisebox{-.3em}{$\widetilde{\;\scriptstyle\MH\to0\;}$}\;
\frac{\alpha}{2\sw^2}\frac{\MW}{\MH}\sigma_{\mathrm{B}},
\eeq
but also more sophisticated approximations are known~\cite{be96a}. For
$\MH\sim60$-$300\GeV$ this effect influences the cross-section about $1\%$
at threshold, but the effect becomes negligible for higher values of
$\MH$ and $E$.

In the large-mass limit~\cite{bo92} for $\MH$ and the top-quark mass
$\Mt$ one gets leading corrections proportional to $\alpha\log\MH$,
$\alpha\log\Mt$, and $\alpha\Mt^2/\MW^2$, where the $\Mt^2$ correction
only enters the isospin part, 
as specified in \refeq{eq:MB}. In the
isospin part, which dominates at low energies, all above-mentioned
leading heavy-mass corrections enter via the renormalization of $\sw$,
i.e.\ are identical to the ones in the well-known quantity $\Delta r$,
and thus can be absorbed by the substitution
$e^2/\sw^2\to4\sqrt{2}\GF\MW^2$. Here $\GF$ is introduced as input
parameter, and $\MW$ is calculated from muon decay. The heavy-mass
correction to the electromagnetic part 
yield contributions of about $0.1\%$ at
LEP2 energies and are thus negligible.
\end{enumerate}
\end{sloppypar}

\subsection{Approximations for radiative corrections}

Taking into account the most important leading corrections, we arrive at
the following improved Born approximation (IBA)
\beqar
\biggl(\frac{d\sigma}{d\Omega}\biggr)_{\mathrm{IBA}} &=& 
\frac{\beta}{64\pi^2s}
\biggl| 2\sqrt{2}\GF\MW^2{\cal M}^\kappa_I\delta_{\kappa-}
+4\pi\alpha(s){\cal M}^\kappa_Q \biggr|^2
\nn\\[.2em] && {}
+\biggl(\frac{d\sigma}{d\Omega}\biggr)_{\mathrm{ISR}}
+\biggl(\frac{d\sigma}{d\Omega}\biggr)_{\mathrm{Coul}}(1-\beta^2)^2,
\label{eq:IBA}
\eeqar
as proposed in \citere{bo92}.
Here the Coulomb correction is accompanied by the weight function 
$(1-\beta^2)^2$, which switches this effect off for $\beta\to1$. 
Figure~\ref{fig:intcs} shows the comparison of the exact $\ord(\alpha)$
correction $\de$ (+~leading higher-order effects via $\De\alpha$ and $\De r$)
to the total unpolarized cross-section
with the IBA for two different choices of the ISR splitting scale $Q^2$.
We recall that the complete bremsstrahlung spectrum is included in $\de$
in this figure.
In \refta{tab:app} we further illustrate the quality of the IBA (with
$Q^2=s$)
by giving its deviation $\De_{\mathrm{IBA}}$ from $\de$ for different
scattering angles $\theta$ using the
soft-photon approximation, i.e.\ excluding hard photons with 
$E_\gamma>\Delta E$ but adding the exact $\Delta E$ terms of the soft-photon
correction to the IBA.
For LEP2 energies we read off an accuracy of 1-2\%. 
Although the IBA remains still valid within $\lsim 5\%$ in the forward
direction, $\theta\lsim 5^\circ$, up energies of $2\TeV$, the IBA is not
able to reproduce the angular distribution of the cross-section in
reasonable approximation for NLC energies.
\btab
\bce
\begin{tabular}{|c|c||r|r|r|r|}
\hline
$\sqrt{s}/\GeV$ & $\theta$ & $\si_{\mathrm{B}}/\fba$ & 
$\De_{\mathrm{IBA}}/\%$ & $\De_{\mathrm{FFA}}/\%$ & $\De_{\mathrm{HEA}}/\%$ \\
\hline\hline
$161$ & $(10^\circ,170^\circ)$ & 
$3753.2$ & $1.5\phantom{0}$ & $0.00$ & $37\phantom{.00}$ \\
\cline{2-6}
&  $10^\circ$ & 
$367.0$ & $1.6\phantom{0}$ & $0.00$ & $36\phantom{.00}$ \\
\cline{2-6}
&  $90^\circ$ & 
$300.7$ & $1.4\phantom{0}$ & $0.00$ & $37\phantom{.00}$ \\
\cline{2-6}
& $170^\circ$ & 
$250.0$ & $1.3\phantom{0}$ & $0.00$ & $37\phantom{.00}$ \\
\hline
$175$ & $(10^\circ,170^\circ)$ & 
$15591$ & $1.3\phantom{0}$ & $0.03$ & $12\phantom{.00}$ \\
\cline{2-6}
&  $10^\circ$ & 
$3380$ &  $1.7\phantom{0}$ & $0.00$ & $10\phantom{.00}$ \\
\cline{2-6}
&  $90^\circ$ & 
$1001$ &  $1.0\phantom{0}$ & $0.05$ & $12\phantom{.00}$ \\
\cline{2-6}
& $170^\circ$ & 
$439$ & $0.59$ & $0.00$ & $12\phantom{.00}$ \\
\hline
$200$ & $(10^\circ,170^\circ)$ & 
$17107$ & $1.5\phantom{0}$ & $0.01$ & $3.7\phantom{0}$ \\
\cline{2-6}
&  $10^\circ$ & 
$6463$ & $1.8\phantom{0}$ & $0.00$ & $2.3\phantom{0}$ \\
\cline{2-6}
&  $90^\circ$ & 
$812$ & $1.4\phantom{0}$ & $0.02$ & $4.7\phantom{0}$ \\
\cline{2-6}
& $170^\circ$ & 
$255$ & $1.3\phantom{0}$ & $0.00$ & $3.8\phantom{0}$ \\
\hline
$500$ & $(10^\circ,170^\circ)$ & 
$4413.1$ & $4.7\phantom{0}$ & $-0.06$ & $-0.85$ \\
\cline{2-6}
&  $10^\circ$ & 
$11604.4$ & $1.9\phantom{0}$ & $0.00$ & $-0.67$ \\
\cline{2-6}
&  $90^\circ$ & 
$75.4$ & $10\phantom{.00}$ & $-0.29$ & $-0.05$ \\
\cline{2-6}
& $170^\circ$ & 
$6.5$ & $14\phantom{.00}$ & $-0.19$ &  $3.5\phantom{0}$ \\
\hline
$1000$ & $(10^\circ,170^\circ)$ & 
$1084.3$ & $11\phantom{.00}$ & $0.06$ & $0.21$ \\
\cline{2-6}
&  $10^\circ$ & 
$3292.3$ & $3.9\phantom{0}$ & $0.00$ & $1.1\phantom{0}$ \\
\cline{2-6}
&  $90^\circ$ & 
$16.7$ & $23\phantom{.00}$ & $0.08$ & $0.54$ \\
\cline{2-6}
& $170^\circ$ & 
$0.6$ & $28\phantom{.00}$ & $-0.77$ & $6.4\phantom{0}$ \\
\hline
$2000$ & $(10^\circ,170^\circ)$ & 
$267.57$ & $22\phantom{.00}$ & $0.12$ & $0.17$ \\
\cline{2-6}
&  $10^\circ$ & 
$823.35$ & $9.7\phantom{0}$ & $0.02$ & $0.64$ \\
\cline{2-6}
&  $90^\circ$ & 
$4.03$ & $39\phantom{.00}$ & $-0.16$ & $0.34$ \\
\cline{2-6}
& $170^\circ$ & 
$0.09$ & $46\phantom{.00}$ & $-2.3\phantom{0}$ & $5.4\phantom{0}$ \\
\hline
\end{tabular}
\ece
\vspace*{-1em}
\caption{
Unpolarized Born cross-section $\si_{\mathrm{B}}$ 
and differences $\Delta$ (normalized to $\si_{\mathrm{B}}$)
between various approximations and exact calculation of the
one-loop RCs for different CMS energies and angles, where
$(10^\circ,170^\circ)$ denotes the angular interval for integrated
cross-sections.
}
\label{tab:app}
\etab

In addition to the IBA \refta{tab:app} also shows a kind of form-factor
approximation (FFA)~\cite{bo92} for the one-loop matrix element:
\beq
\label{eq:FFA}
{\cal M}_{\mathrm{FFA}}^\kappa = 
F_I^\kappa(s,t){\cal M}_I^\kappa + F_Q^\kappa(s,t){\cal M}_Q^\kappa.
\eeq
The four form-factors $F_{I,Q}^\kappa(s,t)$ have been extracted from the exact
one-loop result (which consists of twelve independent ones) by
inspecting the structure of the corrections to the different helicity
amplitudes%
\footnote{A similar form-factor approximation was proposed in
\citere{fl92}, where the structure of anomalous couplings was used as
guideline.}.
Obviously \refeq{eq:FFA} includes all leading and non-leading corrections
associated with the Born structure. The FFA is excellent whenever the
cross-section is sizable. More precisely, in general it approximates the
$\ord(\alpha)$-corrected cross-section within $\sim 0.1\%$, 
and even at the per-cent level if the Born cross-section is suppressed. 
Although this FFA demonstrates that the Born structure indeed
dominates also the one-loop RCs, its evaluation is practically as complicated
as the exact one-loop calculation, which limits its practical use.

For energies above $500\GeV$ a relatively simple approximation of the
virtual and soft-photonic $\ord(\alpha)$ corrections exists~\cite{be93}
which was constructed by a systematic asymptotic expansion for
$E\gg\MW$. The leading corrections are of the form
$\alpha\log(q_1^2/\MW^2)\log(q_2^2/\MW^2)$ with $q_i^2=s,t,u$ and
originate from vertex and box graphs. Depending on the hierarchy of
scales the heavy-mass effects in longitudinal W-boson production can be
additionally enhanced: for $s\gg\Mt^2,\MH^2\gg\MW^2$ power corrections
of $\ord(\alpha\Mt^2/\MW^2)$ and $\ord(\alpha\MH^2/\MW^2)$ show up
(incomplete screening); for $\Mt^2,\MH^2\gg s\gg\MW^2$ one observes
so-called delayed unitarity effects leading to terms of 
$\ord(\alpha s/\MW^2\log\Mt)$ and 
$\ord(\alpha s/\MW^2\log\MH)$. By construction the high-energy
approximation (HEA) asymptotically approaches the exact one-loop RCs, but
already for energies of 500-1000$\GeV$ the deviation is only of the
order of 1\% for small and intermediate scattering angles, 
as can be read off from \refta{tab:app}.

\section{Four-fermion production}
\label{se:4fermion}

\subsection{Born cross-section and its complications}

The simplest extension of on-shell W-pair production to the off-shell 
case consists in the convolution 
\beq
\si(s) = B_{f_1\bar f_2}B_{f_3\bar f_4}
\int dk_+^2\;\rhow(k_+^2) 
\int dk_-^2\;\rhow(k_-^2)
\si_{\Pe\Pe\PW\PW}(s,k_+^2,k_-^2),
\eeq
of the ``off-shell cross-section'' $\si_{\Pe\Pe\PW\PW}$
with the Breit-Wigner distribution
\beq
\rhow(k_\pm^2) = \frac{1}{\pi} \frac{\MW\Ga_\PW}
{\left|k_\pm^2-\MW^2+i\MW\Ga_\PW\right|^2}
\;\;\raisebox{-.3em}{$\stackrel{\displaystyle\longrightarrow}
{\scriptstyle\Ga_\PW\to0}$}\;\;\de(k_\pm^2-\MW^2),
\eeq
as proposed in \citere{Mu86}. Here $\Ga_\PW$ denotes the
decay width of the W boson, and $B_{f_i\bar f_j}$ the
branching ratio for $\PW\to f_i\bar f_j$. 
For $\Ga_\PW\to0$ one clearly obtains the narrow-width approximation.
Obviously this approach does not lead to gauge-invariant results for two
reasons: Firstly, not all diagrams for the process 
$\Pep\Pem\to f_1\bar f_2 f_3\bar f_4$ are taken into account, but only the
doubly resonant ones which factorize into production and decay.
Secondly, the inclusion of finite-width effects goes beyond a pure
perturbative expansion and deserves particular care.

The first problem can in principle be solved by taking into account
all tree graphs contributing to $\Pep\Pem\to f_1\bar f_2 f_3\bar f_4$
which includes in addition so-called ``background diagrams'' with only
one or even no resonant W boson. Figure~\ref{fig:bgdiag} shows some typical
topologies for such graphs at tree level.
\begin{figure}
\begin{center}
\begin{picture}(10.5,2.0)
\put(-5.5,-23.2){\includegraphics{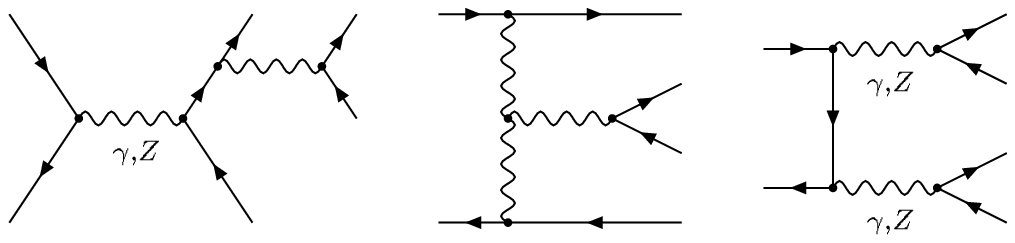}}
\end{picture}
\end{center}
\caption{Typical topologies for background graphs to 
\protect{$\Pep\Pem\to\PW\PW\to f_1\bar f_2 f_3\bar f_4$} at tree level.}
\label{fig:bgdiag}
\efi
The number of background diagrams depends on the final state, ranging
from 6 for $\Pep\Pem\to\mu^-\bar\nu_\mu\tau^+\nu_\tau$ to 53 for
$\Pep\Pem\to\Pep\nu_\Pe\Pem\bar\nu_\Pe$. The numerical impact of these
graphs in general is of ${\cal O}(\Ga_\PW/\MW)\sim 3\%$ relative to the 
doubly resonant contributions and can be reduced by invariant-mass cuts
\beq
\MW-\De_\PW < M_{f_1\bar f_2},M_{f_3\bar f_4} < \MW+\De_\PW
\eeq
by a factor of $\De_\PW/\MW$, where $M_{f_i\bar f_j}$ is the invariant
mass of the corresponding fermion pair.
However, in this context it should be mentioned that the influence of 
the background diagrams can be enhanced in
certain regions of phase space such as for forward $\Pe^\pm$ scattering.
The numerical evaluation of cross-sections including this background has
been performed by several authors using Monte-Carlo or semianalytical
methods and is reviewed elsewhere~\cite{oh96} 
(see also \citere{CERN9601mcgen}). 

The second problem mentioned above is seen by inspecting the general
structure of the amplitude,
\beq
\label{eq:amp}
{\cal M} = \frac{R_{+-}(k_+^2,k_-^2)}{(k_+^2-\MW^2)(k_-^2-\MW^2)}
+\frac{R_{+}(k_+^2,k_-^2)}{k_+^2-\MW^2}
+\frac{R_{-}(k_+^2,k_-^2)}{k_-^2-\MW^2}+N(k_+^2,k_-^2),
\eeq 
which contains the doubly, singly, and non-resonant contributions
$R_{+-}$, $R_\pm$, and $N$, respectively. Gauge invariance implies that
${\cal M}$ does not depend on the gauge-fixing procedure of the
perturbative calculation (usually quantified by a gauge parameter)
and that ${\cal M}$
obeys Ward identities which guarantee gauge cancellations
and thus unitarity. However, this is only the case for the sum but not
for the single terms in \refeq{eq:amp}. Therefore, introducing a 
(possibly running) finite width via
\beq
\frac{1}{k_\pm^2-\MW^2} 
\;\;\to\;\;\frac{1}{k_\pm^2-\MW^2+i\MW\Ga_\PW(k_\pm^2)}
\eeq
breaks gauge invariance by terms which are formally of 
${\cal O}(\Ga_\PW/\MW)$ \cite{ae93}. However, these
gauge-invariance-breaking terms can be further enhanced in the presence
of small scales or if gauge cancellations are disturbed.
\begin{figure}
\begin{center}
\begin{picture}(12.5,2.5)
\put(-4.5,-22.7){\includegraphics{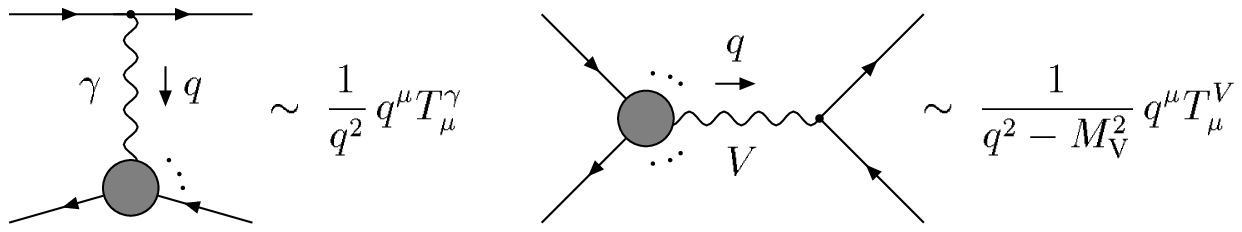}}
\put(1.8,-.6){\reffi{fig:gcanc}a)}
\put(8.6,-.6){\reffi{fig:gcanc}b)}
\end{picture}
\end{center}
\caption{Typical situations for gauge cancellations for
a) \protect{$|q^2|\ll s$} due to electromagnetic gauge invariance, and for
b) \protect{$q^0\gg \MV$} due to SU(2) gauge invariance.}
\label{fig:gcanc}
\efi
Gauge cancellations are required for a well-behaved matrix element if
external fermionic currents $\bar u(p)\ga^\mu u'(p')$ become
proportional to the momentum $q^\mu$ of the gauge boson which directly
couples to this current.
Figure~\ref{fig:gcanc} shows typical situations where gauge cancellations
are relevant:
Figure~\ref{fig:gcanc}a) illustrates that $q^\mu T^\ga_\mu$ must
compensate the pole $1/q^2$ in the photon propagator for forward electron 
scattering ($|q^2|\ll s$).
Figure~\ref{fig:gcanc}b) shows
the production of an effectively longitudinally polarized massive gauge
boson \PV\ at high energies ($q^0\gg \MV$) where cancellations in 
$q^\mu T^V_\mu$ are necessary in order to guarantee unitarity. These
cancellations are ruled by the Ward identities
\beq
\label{eq:WIs}
q^\mu T^\ga_\mu     = 0, \qquad
q^\mu T^Z_\mu       = i\MZ T^\chi, \qquad
q^\mu T^{W^\pm}_\mu = \pm\MW T^{\phi^\pm},
\eeq
where the first one expresses electromagnetic current conservation,
and the others imply the Goldstone-boson equivalence theorem, which
relates the amplitudes for Z and W bosons to the ones for their
respective would-be Goldstone bosons $\chi$ and $\phi$.

In the literature several methods for the introduction of finite decay
widths in amplitudes were proposed which are discussed in view of the
above gauge-invariance issue elsewhere~\cite{be96b}. Here we just mention
the field-theoretically most convincing approach of the so-called
fermion-loop scheme~\cite{flscheme}, which consists of the consequent
inclusion of all fermionic one-loop corrections (self-energy, vertex and
box corrections) to tree-level amplitudes using Dyson-resummed
propagators. This approach is manifestly gauge-parameter-independent, 
preserves the crucial Ward identities \refeq{eq:WIs}, and introduces the
complete finite-width effects of unstable particles that decay exclusively 
into fermion pairs, such as the weak gauge bosons Z and W.

\subsection{Radiative corrections}

Owing to its complexity a complete calculation of the $\ord(\alpha)$ RCs
to four-fermion production has not been presented yet, and hopefully is
not needed. So far only leading RCs have been included in theoretical
predictions.

For on-shell W-pair production we have seen in the last section that
leading-log ISR, leading electroweak corrections in $\GF$ and
$\alpha(q^2)$, and Coulomb correction approximate the complete
$\ord(\alpha)$-corrected cross-section within 1-2\% at LEP2 energies.
Therefore we expect that this is also the typical theoretical
uncertainty in the corresponding leading-log improved predictions for
four-fermion production, at least in regions of phase space where the
doubly resonant graphs are dominant. In practice ISR is treated either
using the structure-function approach analogously to \refeq{eq:ISR} or
using parton-shower methods (see e.g.\ 
\citeres{CERN9601wwcross,CERN9601mcgen} for details). Moreover, the
absorption of the leading weak corrections arising from renormalization
($\Delta\alpha$, $\Delta r$) follows exactly the same lines as in the
on-shell case. However, the Coulomb correction is considerably modified
by finite-width effects~\cite{fa93}. The singular $\beta^{-1}$ correction
factor to the on-shell cross-section is regularized in the off-shell
case:
\beq
\delta\sigma_{\mathrm{Coul}} = \sigma_{\mathrm{B,2-res}} 
\frac{\alpha\pi}{2\bar\beta}
\biggl[ 1-\frac{2}{\pi}\mathrm{arctan}
\biggl(\frac{|\beta_M|^2-\bar\beta^2}{2\bar\beta\Im\beta_M}\biggr)
\biggr],
\eeq
with $\sigma_{\mathrm{B,2-res}}$ denoting the doubly resonant part of
the Born cross-section, and
\beqar
\bar\beta &=& \sqrt{1-2(k_+^2+k_-^2)/s+(k_+^2-k_-^2)^2/s^2},
\nn\\
\beta_M &=& \sqrt{1-4(\MW^2-i\MW\Gamma_\PW)/s}.
\eeqar
The on-shell singularity is recovered by taking $\Gamma_\PW\to0$ after
setting $\bar\beta^2=\Re\beta_M^2=\beta^2$. For finite $\Gamma_\PW$ the
$\bar\beta^{-1}$ correction is screened if the average velocity
$\bar\beta$ becomes smaller than $|\beta_M|\gsim\sqrt{\Gamma_\PW/\MW}$.
The maximal effect of $\sim 6\%$ is reached at threshold, while this
correction still amounts to 2-3\% for CMS energies around 
$\sqrt{s}\sim 175\GeV$.

The approximation of RCs that includes only the leading effects
described above is certainly not able to
match the aimed experimental $1\%$ accuracy. Therefore at least the
$\ord(\alpha)$ corrections to the doubly resonant contributions have to
be taken into account. A first step towards this direction is reached if
finite decay widths are introduced via the fermion-loop scheme, as
briefly described in the previous subsection. In this approach the
complete fermion-loop corrections of $\ord(\alpha)$ are included which
represent a substantial part of the complete $\ord(\alpha)$ RCs. Although
there is a natural generalization of the gauge-invariant fermion-loop
resummation to the complete (fermionic + bosonic) RCs within the
background-field method~\cite{de95}, the general problem is not yet
under control. On the one hand, gauge invariance, which guarantees the
Ward identities \refeq{eq:WIs}, requires the inclusion of all (doubly,
singly, and non-resonant) $\ord(\alpha)$ RCs; on the other hand, a
gauge-parameter dependence remains at the loop order that is incomplete,
e.g.\ in $\ord(\alpha^2)$ in a complete Dyson-resummed one-loop
calculation. This direction certainly deserves further investigations.

\begin{sloppypar}
For a first approach to include also bosonic $\ord(\alpha)$ RCs one can
perform a pole expansion~\cite{ve63} about the double resonance and keep
only the doubly resonant terms, which are known to be
gauge-independent. 
Although the reliability of this approach near threshold is not clear, 
the relative uncertainty
for energies of $\sqrt{s}\gsim 175\GeV$ is expected to be of order
$(\alpha/\pi)(\Gamma_\PW/\MW)\ord(1)\lsim 0.1\%$. Technically the
pole-scheme calculation consists in taking only the contribution of
$R_{+-}(k_+^2,k_-^2)$ in the matrix element, as specified in 
\refeq{eq:amp}, by setting $k_\pm^2=\MW^2$ in $R_{+-}$. This procedure
considerably reduces the number of relevant diagrams but includes the
complete $\ord(\alpha)$ RCs to the on-shell production and the subsequent
decay of the W~bosons. Additionally doubly resonant, but
non-factorizable diagrams contribute, which arise from photonic
initial-final and final-final state interactions, as indicated in 
\reffi{fig:phdiag}. 
\begin{figure}
\begin{center}
\begin{picture}(12.5,2.5)
\put(-4.5,-22.7){\includegraphics{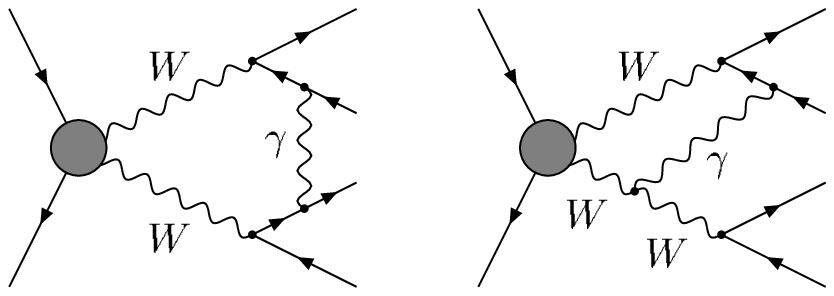}}
\end{picture}
\end{center}
\caption{Examples for non-factorizable photonic corrections.}
\label{fig:phdiag}
\efi
It turns out that only infrared photons are relevant
for the doubly resonant part. In a special kind of
soft-photon approximation it was shown in \citere{fa94} that the
doubly resonant virtual + soft-photonic $\ord(\alpha)$ contributions to
the total cross-section cancel. In the same approximation this kind of
correction was calculated in \citere{me96} for the differential
cross-section, where non-vanishing corrections to the invariant-mass
distributions occur. Unfortunately a complete numerical evaluation of
$\Pep\Pem\to\PW\PW\to 4f$ in the pole scheme has not yet been presented
in the literature.
\end{sloppypar}

\section{Conclusions}
\label{se:concl}

We have reviewed theoretical aspects of W-pair production in $\Pep\Pem$
annihilation by placing the emphasis on $\ord(\alpha)$ RCs. The dominant
$\ord(\alpha)$ RCs are of universal origin, viz.\ collinear ISR,
renormalization effects due to $\Delta\alpha$ and $\Delta r$, and the
Coulomb singularity near threshold. They are also known for
the four-fermion production process $\Pep\Pem\to 4f$ and thus can be 
easily included in present-day Monte-Carlo event generators. However, a
comparison of these dominant RCs with exact $\ord(\alpha)$ calculations
for on-shell W-pair production reveals that for typical LEP2 energies
one cannot expect to approximate the full $\ord(\alpha)$-corrected
cross-section for $\Pep\Pem\to 4f$ to better than 1-2\% by including
only these leading effects. Thus, also non-leading $\ord(\alpha)$
corrections have to be taken into account to match the aimed
experimental error of roughly 1\% at LEP2.

The introduction of finite decay widths of resonating particles in
amplitudes is non-trivial owing to problems with gauge invariance. At
tree level a convincing solution exists for gauge-boson resonances,
however, the general problem
requires further investigations. A
first approximation for the inclusion of $\ord(\alpha)$ RCs to
$\Pep\Pem\to\PW\PW\to 4f$ can be obtained via a pole expansion about the
double resonance and keeping only the doubly resonant contributions.
This procedure includes all $\ord(\alpha)$ RCs to the on-shell W-pair 
production and the W~decay, but also some non-factorizable photonic
corrections arising from initial-final and final-final state
interactions.

In this short review we could not consider final-state interactions
which are associated with hadronization, such as color reconnection and
Bose-Einstein correlations, which both influence the W-boson mass
determination at LEP2. For these effects we refer to the literature (see 
\citere{CERN9601wmass} and references therein).

\section*{Acknowledgements}
The author would like to thank the organizers for the kind invitation 
and for providing a very warm atmosphere during the conference.
A.~Denner is also gratefully acknowledged for helpful
discussions.

\end{document}